\documentclass[aps,pra,twocolumn,longbibliography,groupedaddress,10pt]{revtex4-1}
\usepackage{graphicx}
\usepackage{multirow}
\usepackage{amsmath}
\usepackage{amssymb}
\usepackage[utf8]{inputenc}
\usepackage{array}
\usepackage{xcolor}
\definecolor{dark-green}{RGB}{0, 128, 0}
\definecolor{dark-red}{rgb}{0.4,0.15,0.15}	
\definecolor{dark-blue}{rgb}{0.15,0.15,0.4}	
\definecolor{medium-blue}{rgb}{0,0,0.5}		

\usepackage[caption=false]{subfig}

\usepackage{hyperref}
\hypersetup{colorlinks, linkcolor={dark-red}, citecolor={dark-blue}, urlcolor={medium-blue}}

\newcommand{\Esc}{E_{\rm sc}}
\newcommand{\Eext}{E_{\rm ext}}
\newcommand{\bEsc}{\mathbf{E}_{\rm sc}}
\newcommand{\bEext}{\mathbf{E}_{\rm ext}}
\newcommand{\Isat}{I_{\rm sat}}
\newcommand{\bnabla}{\boldsymbol\nabla}

\begin{document}

\title{Anisotropic nonlinear refractive index measurement \\ of a photorefractive crystal via spatial self-phase modulation}

\author{Omar Boughdad}
\thanks{Both authors contributed equally}
\author{Aur\'elien Eloy}
\thanks{Both authors contributed equally}
\author{Fabrice Mortessagne}
\author{Matthieu Bellec}
\email{matthieu.bellec@inphyni.cnrs.fr}
\author{Claire Michel}
\email{claire.michel@univ-cotedazur.fr}

\affiliation{
Universit\'e C\^ote d'Azur, CNRS, Institut de Physique de Nice, France}


\begin{abstract}
We show that the refractive index modification photoinduced in a biased nonlinear photorefractive crystal can be accurately measured and controlled by means of a background incoherent illumination and an external electric field.
The proposed easy-to-implement method is based on the far-field measurement of the diffraction patterns of a laser beam propagating through a self-defocusing medium undergoing spatial self-phase modulation.
For various experimental conditions, both saturation intensity and maximum refractive index modification have been measured.
We also clearly evidence and characterise the anisotropic nonlinear response of the crystal in the stationary regime.
\end{abstract}


\maketitle



\section{Introduction}

In the past few decades, photorefractive (PR) crystals \cite{yeh1993} have been extensively used as a suitable medium for nonlinear optics applications \cite{gunter2007}.
With a controllable bias voltage and under incoherent illumination, the photoinduced optical properties of PR crystals can indeed be efficiently engineered.
Numerous nonlinear optics experiments have been carried out, mainly focused on the study of the linear and nonlinear light transport properties.
On the one hand, non-diffracting optical beams allow to fabricate non-permanent photonic lattices, i.e. arrays of coupled waveguides.
It has been widely used to observe various families of optical discrete solitons \cite{Fle03, Fle05, Led08}, to study wave properties of quasicrystals \cite{Bog16} and graphene-like structures \cite{Pel07, Son19} as well as to evidence weak or strong transverse localisation in disordered lattices \cite{schwartz2007, boguslawski2017}.
On the other hand, in homogeneous configuration, nonlinear patterns formation dynamics \cite{Den98, Mar10, Cau12} as well as quantum hydrodynamical features of light \cite{wan2007, sun2012, michel2018} have been investigated.

In order to get better insights on such light transport properties and allow quantitative comparison with theory and numerics, a key experimental parameter to monitor is the amplitude of the photoinduced refractive index modification.
In principle, this value can be accurately controlled through an external voltage and an incoherent illumination but, surprisingly, is roughly estimated in most experiments (typically of the order of $10^{-4}$).
Moreover, when measured, the calibration method employed and the associated uncertainties are rarely addressed.
Recently, Armijo \textit{et al.} proposed an absolute calibration method to measure the refractive index of photoinduced 1D and 2D photonic lattices \cite{armijo2014}.
These measurements are very consistent with theory but specific to lattice configurations in a focusing, highly saturated regime.
In a previous work, we also addressed a method based on the displacement of a weak optical defect in a strong optical gaussian environment to extract the maximum refractive index and the saturation intensity \cite{michel2018}.
The measurements were in quite good accordance with the expected theoretical values but not easy to implement and time consuming.

\begin{figure}[t]
	\centering
	\includegraphics{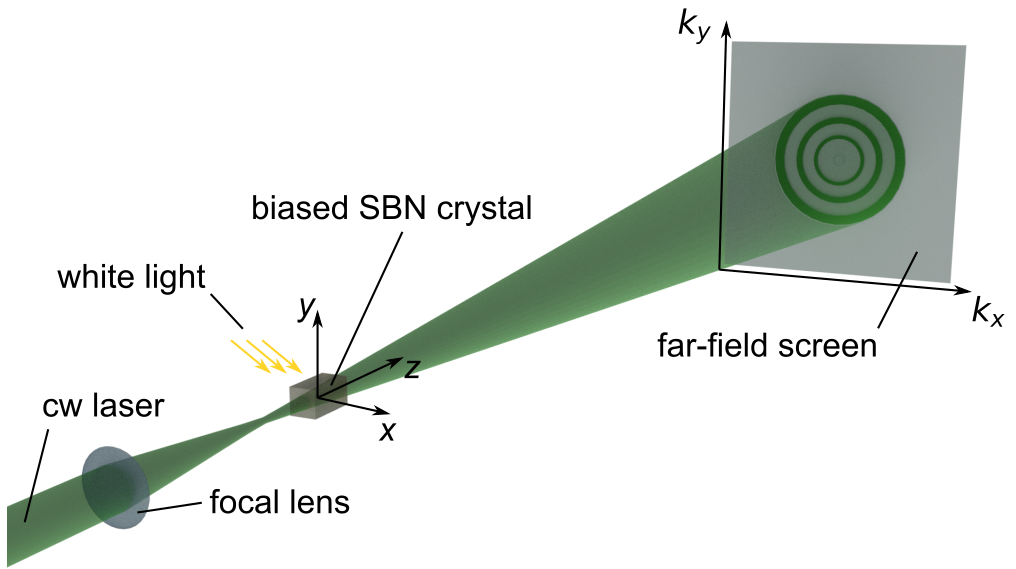}
	\caption{Sketch of the experimental setup. A focused cw laser beam propagates in a biased SBN photorefractive crystal.
	The beam is linearly polarised along $x$, corresponding to the $c$-axis.
	The crystal is thinner than the Rayleigh length and placed out of focus which garanties a quasi-constant beam curvature radius.
	The spatial (2D) spectrum is measured in the far-field.
	The external voltage (resp. the white light illumination) controls the maximum value of the photoinduced refractive index modification $\Delta n_{\rm max}$ (resp. the saturation intensity $I_{\rm sat}$).}
	\label{fig:fig-exp}
\end{figure}

When an ultrashort pulse travels in a nonlinear medium new frequencies appear within its spectrum as the input intensity increases, via the self-phase modulation effect \cite{Agr12, Fin18}.
Each new frequency is associated to a 2$\pi$-phase shift and thus to a given value of the nonlinear refractive index.
Similarly, when a cw spatially limited wavepacket propagates in a nonlinear medium, new \textit{spatial} frequencies are generated as the input intensity increases.
In this case, it simply consists of rings appearing in the far-field, as illustrated in Fig.~\ref{fig:fig-exp}.
This approach already allowed to measure cw nonlinear effects in liquid crystals \cite{durbin1981, Dip18}, graphene suspensions \cite{Sha19} as well as hot atomic vapours \cite{San18, fontaine2018} to cite a few.

Here, we exploit spatial self-phase-modulation (sSPM) to measure the nonlinear photorefractive response of a Strontium Barium Niobate (SBN) crystal.
Both saturation intensity and maximum refractive index modification have been measured with respect to the input intensity, external voltage and incoherent illumination amplitude.
We also clearly evidenced and measured the anisotropic nonlinear response of the crystal, in good agreement with reported measurement in lattices \cite{armijo2014}.

The paper is organised as follow.
We first briefly describe the nonlinear response of the PR crystal, and the nonlinear propagation of a laser beam polarised along its crystallographic axis.
Then we present the sSPM effect, which describes the nonlinear interference pattern formed in the far-field at the output of the nonlinear medium.
This effect results from the nonlinear phase accumulated by the laser beam along its propagation.
Then we show how this approach allows to measure the absolute nonlinear refractive optical index of the crystal. Experimental and numerical results are compared.

\section{Photorefractive model and nonlinear CW laser propagation}

In this section, we shortly describe the nonlinear photorefractive effect \cite{denz2003} occuring when a photorefractive crystal is illuminated by a laser beam.
In the regions of high intensity, impurities charge carriers go from the valence to the conduction band of the crystal and become mobile.
They therefore move to different locations in the crystal, and two transport mechanisms are considered.
First, charges can be diffused toward a lower charge concentration place.
Second, they can be drifted whether by the internal electric field generated by the photovoltaic effect, and/or by an external electric field.
The driven motion of the charge carriers occurs in a preferential direction, which is naturally determined by the direction of the polar axis ($c$-axis) of the crystal.
An exeternal electric field applied along the $c$-axis reinforces the directed motion of the charges.
Finally, the electrons (or holes) recombine with free sites of the crystal lattice, creating a new distribution of charges, resulting in a spatially distributed space charge electric field.
This field modulates the optical refractive index of the crystal, through the linear electro-optic Pockels effect.
Depending on the orientation of the external electric field with respect to the $c$-axis, the refractive index change can be either positive (focusing effect) or negative (defocusing effect).
For a linear polarisation in the transverse $(x,y)$ plane, parallel to the crystallographic $c$-axis, $z$ being the direction of propagation, the variation of the refractive index in the transverse plane is then given by:
\begin{equation}
	\Delta n(I) = -\frac{1}{2}n_0^3r_{33}E_{\rm sc}[I(x,y)],
	\label{eq:Deltan}
\end{equation}
where $n_0=2.36$ in the linear refractive index of the bulk crystal along the $c$-axis.
In this case, the electro-optic coefficient is $r_{33}=250$ pm/V and $E_{\rm sc} = |\mathbf{E}_{\rm sc}(I)|$ is the amplitude of the space charge electric field which depends on the intensity $I(x,y)$ of the optical beam.

\subsection{Anisotropic description}

In a biased photorefractive crystal with an external electric field $\bEext$ applied along the $c$-axis, and in the framework of the band-transport model proposed by Kukhtarev \textit{et al.} \cite{kukhtarev1978}, the space charge electric field can be expressed in terms of an electrostatic potential $\phi$.
It consists in a light induced potential $\phi_0$ and an external bias term $-|\bEext|x$, considering the $c$-axis aligned with the $x$-axis.
This results in an anisotropic description of the photorefractive effect, leading to $\bEsc=\bEext-\bnabla\phi_0$ \cite{zozulya1995}.
The quantity $\bEsc-\bEext=-\bnabla\phi_0$ is often referred as a \textsl{screening field} within the crystal, generated by the light induced potential \cite{diebel2014}.
Considering that the system is in a steady state, that the photovoltaic effect is negligible, and that the drift effect overcomes diffusion for the charge carriers migration, the potential equation for the electrostatic field $\phi_0$ is given by
\begin{equation}
	\bnabla_\perp^2\phi_0 + \bnabla_\perp\ln{(1+\tilde I)}\cdot\bnabla_\perp\phi_0=|\bEext|\partial_x\ln{(1+\tilde I)},
	\label{eq:potential}
\end{equation}
where $\bnabla_\perp$ is the gradient in the transverse plane in cartesian coordinates.
$\tilde I=I/\Isat$ is the laser beam intensity normalised to the saturation intensity (often associated to a dark intensity), which accounts for the ratio between the thermal and the photoinduced excitations.
In experiments, an incoherent white light illumination is usually used as a background to artificially create a contribution to the thermal excitation in a controllable way.

\subsection{Isotropic approximation}

If the spatial dimension of the optical beam is much larger than the spatial scale of the electric field, the system adopts a (1+1) dimensional geometry, with a quasi-homogeneous illumination of the crystal \cite{denz2003}.
In this case, the description of the photorefractive effect is slightly different than the previous case and leads to an isotropic 1D description of the phenomenon \cite{christodoulides1995}.
The space charge electric field thus comes to a simpler expression, $\Esc = \Eext/(1+\tilde I)$.
In this model, the refractive index variation reads:
\begin{equation}
	\Delta n(I) = -\frac{1}{2}n_0^3r_{33}\Eext\frac{1}{1+\tilde I}.
	\label{eq:Deltan_iso}
\end{equation}
In the latter expression, the light intensity induces a \textsl{screening} of the refractive index variation.
Indeed, when the laser intensity is zero, the absolute refractive index variation $|\Delta n|$ is maximised.
On the contrary, this variation tends to zero when $\tilde I$ becomes much greater than unity.
The screening effect saturates for high enough laser intensities, and the level of saturation is controlled by the parameter $\Isat$.
Although not strictly valid in the (2+1)D configuration, this isotropic description will be considered for the analysis of the experimental results.

\subsection{Nonlinear cw laser propagation}

We model the propagation of a laser beam of amplitude $E(x,y,z)$ and intensity $I=|E|^2$, polarised along the $c$-axis of the crystal with the nonlinear Schr\"odinger equation (NLSE):
\begin{equation}
	i\partial_zE = -\frac{1}{2k_0n_0}\bnabla_\perp^2 E - i\alpha E + k_0\Delta n(I) E,
	\label{eq:NLS}
\end{equation}
where $z$ is the propagation direction, $k_0=2\pi/\lambda_0$ with $\lambda_0$ the vacuum wavelength of the laser beam, $n_0$ the base index of refraction of the bulk crystal, $\Delta n(I)$ the nonlinear variation of the refractive index described by equations (\ref{eq:Deltan}) to (\ref{eq:Deltan_iso}), and $\alpha$ the linear absorption coefficient of the bulk material.
The absorption coefficient has been measured experimentally and is estimated at $\alpha=1.319$ cm$^{-1}$ at 532 nm.

Applying a change of variable, that is $E~=~A\cdot \mathrm{e}^{ik_0\Delta n_{\rm max} z}$ with $\Delta n_{\rm max}=\frac{1}{2}n_0^3r_{33}\Eext$ leads to a new propagation equation for $A$ :
\begin{equation}
	i\partial_z A = -\frac{1}{2k_0n_0}\bnabla_\perp^2 A - i\alpha A + k_0\Delta n_{\rm max}\left(\frac{\tilde I}{1+\tilde I} \right)A.
	\label{eq:NLS_A}
\end{equation}
This corresponds to a change of reference for the bulk refractive index, and defines a new expression for the nonlinear variation of the refractive index that reads :
\begin{equation}
	\Delta n(I) = \frac{1}{2}n_0^3r_{33}\Eext\frac{\tilde I}{1+\tilde I} = \Delta n_{\rm max}\frac{\tilde I}{1+\tilde I},
	\label{eq:Deltan_sat}
\end{equation}
in which the saturation of $\Delta n$ with respect to the laser intensity appears clearly.
The coefficient $\Delta n_{\rm max}$ thus corresponds to the maximum variation of the nonlinear refractive index so that $\Delta n_{\rm max} = \frac{1}{2}n_0^3r_{33}\Eext$.


\section{Spatial self-phase modulation and nonlinear refractive index measurement}

In order to fully characterise the nonlinear index variation induced by a laser beam, we take advantage of the spatial self-phase modulation (sSPM) to measure the accumulated nonlinear phase shift of a Gaussian beam propagating through the SBN crystal.
sSPM is a nonlinear effect which refers to the modulation of the phase in the transverse plane of a laser beam propagating in a nonlinear medium.
This phase modulation is proportionnal to the transverse intensity profile of the laser beam.
Given the relation~\eqref{eq:Deltan_sat}, the variation of the intensity in the transverse plane as well as along the propagation results in a nonlinear phase shift $\Delta \Psi_{\text{NL}}(x,y)$ that can be written as:

\begin{equation}
 \Delta \Psi_{\text{NL}}(x,y) = k_0\int_{0}^{L_{z}} \Delta n(r,z)\,dz,
    \label{eq:DeltaPhiNL}
\end{equation}
\noindent
with $L_{z}$ the length of the photorefractive crystal.
For a laser beam with a transverse Gaussian profile of radius at $1/{\rm e}^2$ denoted $w_p$, one can show that, in a non-saturated regime,
\begin{equation}
	\Delta \Psi_{\text{NL}}(x,y)  = \Delta \Psi_0 \exp{\left(-r^2/w_p^2\right)},
\end{equation}
$r$ being the radial coordinate such as $\mathbf{r} = (x,y)$.
Consequently, for each point $r_1$ on the curve of $\Delta \Psi_{\text{NL}}$, it is always possible to find a second point, $r_2$, having the same slope.
Hence, fields coming from regions close to $r_1$ and $r_2$ present the same wavevector and two-wave interferences are expected.
If $\Delta \Psi_0 \gg 2\pi$, the diffraction pattern the far-field is composed of a set of concentric rings.
Bright or dark rings, that is to say maximum constructive (resp. destructive) interference, occur when $\Delta \Psi_{\text{NL}}(r_1) - \Delta \Psi_{\text{NL}}(r_2) = m\pi$, with $m$ an even (resp. odd) integer.
Finally, the number of rings $N_{\mathrm{rings}}$ can be estimated thanks to the relation~\cite{durbin1981}:

\begin{equation}
 N_{\mathrm{rings}} \simeq \frac{\Delta \Psi_0}{2\pi}.
\end{equation}
In the following, we present the use of sSPM to access experimentally the nonlinear refractive index of SBN by counting the rings appearing  in the diffraction pattern in the far-field.


\section{Experimental setup}

The experimental set-up is depicted in Fig.~\ref{fig:fig-exp}.
The nonlinear medium is a $5\times5\times20\,\mathrm{mm}^3$ SBN:61 crystal, doped with cerium (0.002 \%) to enhance its photoconductivity.
The crystal is shone by a laser beam (wavelength $\lambda_0 = 532\,\mathrm{nm}$) with a Gaussian transverse profile whose radius at $1/\mathrm{e}^2$ is $220\,\mu\mathrm{m}$ at the entrance of the PR crystal, propagating along the $z$ axis.
The position of the crystal along $z$ is fixed in order to keep constant the incident radius of curvature of the wavefronts estimated to be $21\,\mathrm{cm}$ on the entrance facet, and considered constant along the propagation of the light through the crystal.
The polarisation of the light is maintained horizontal (i.e. along $x$), and parallel to the $c$-axis to maximise the electrooptic coefficient $r_{33}$ of SBN.
Two electrodes allow to create an external electric field $\Eext$ along the $c$-axis.
Changing $\Eext$ results in a modification of the maximal nonlinear refractive index $\Delta n_{\mathrm{max}}$, as indicated in Eq.\eqref{eq:Deltan_iso}.
The dynamics of the saturation of the photorefractive effect is controlled by illuminating the crystal with an incoherent white light, providing a control over $\Isat$ and so the dark intensity.
As explained previously, the white light enhances thermal effects that increase the charge carriers' mobility, leading to an easier recombination process.
The presence of white light thus increases $\Isat$, hence one needs a higher laser intensity to saturate the refractive index variation.
The illumination remains unmoved over all experiments, only its power is varied, in order to ensure the best possible control over the variation of $\Isat$. 
The maximum accessible white light power is estimated to be $3\,\mathrm{W}$, provided by a halogen lamp of nominal power $60\,\mathrm{W}$.
Practically, the nonlinearity is thus controlled by changing the intensity $I$ of the laser, the electric field $\Eext$ and the saturation intensity $I_{\mathrm{sat}}$.
Finally, the Fourier transform of the outgoing light is imaged on a sCMOS camera placed one meter away from the output facet. This distance is large enough to considerer being in the far-field.

\begin{figure}[t]
	\centering
	\includegraphics{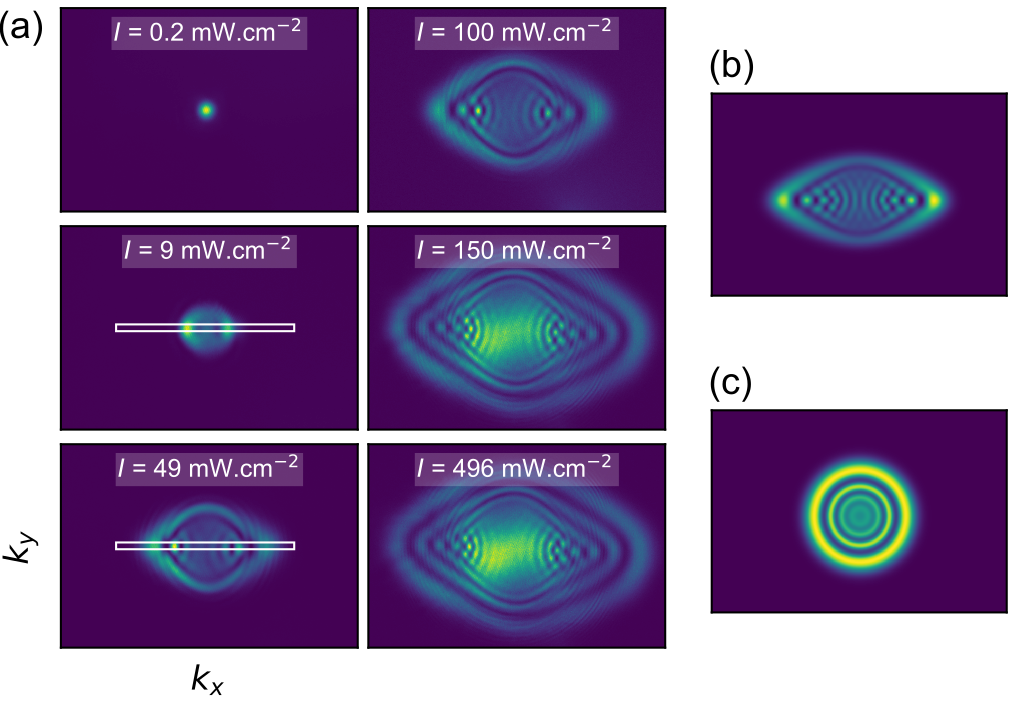}
	\caption{Intensity distribution of the spatial spectrum measured in the far-field, i.e. ($k_x, k_y$) plane, for different initial laser intensities.
	(a) Experimental measurements.
	(b) Numerical image obtained for $I = 100$ mW.cm$^{-2}$, by solving the coupled Eqs.~(\ref{eq:NLS}) and~(\ref{eq:potential}).
    (c) Numerical image obtained for the same optical intensity by solving the Eq.~(\ref{eq:NLS}) in the isotropic approximation.
	For each panel, $k_x$ (resp. $k_y$) spans from -52.5 mm$^{-1}$ to 52.5 mm$^{-1}$ (resp. -36 mm$^{-1}$ to 36 mm$^{-1}$).}
	\label{fig:images}
\end{figure}


\section{Results and discussion}

\begin{figure}[t]
  \centering
  \includegraphics[width=8.6cm]{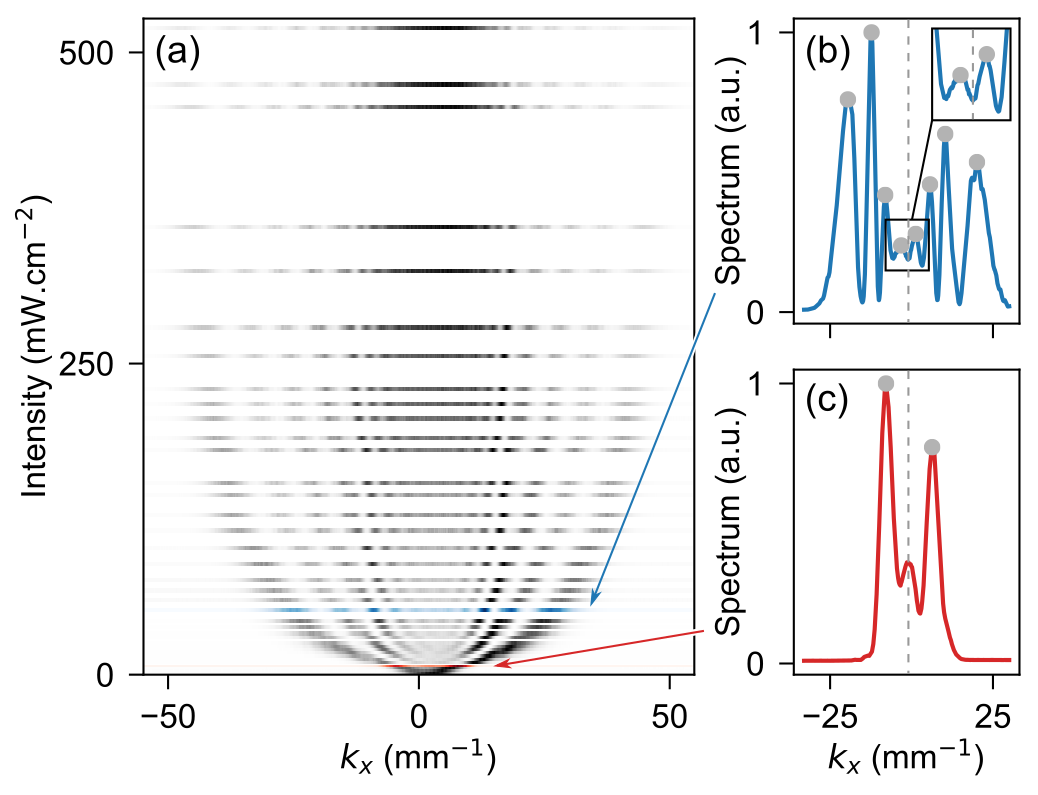}
  \caption{(a) Experimental spectrum profiles (cut along $k_x$) for various input intensities.
  Black (resp. white) corresponds to the maximum (resp. minimum) spectral power which is normalised to 1 for each measurement.
  The widths of the profiles have been artificially increased for clarity.
  (b-c) Typical profiles at $I=9$ mW.cm$^{-2}$ and $I=49$ mW.cm$^{-2}$ (red and blue curves respectively).
  The gray points count twice the number of rings $N_{\rm ring}$.
  The concave (resp. convexe) behaviour at $k_x=0$ (dashed gray line) in the red (resp. blue) curve defines the nonlinear phase shift as $\Delta \Psi_0 = 2\pi{}N_{\mathrm{rings}}$ [resp. $\Delta \Psi_0 = (2N_{\mathrm{rings}} - 1)\pi$], see text for details.}
  \label{fig:profiles}
\end{figure}

Figure \ref{fig:images}(a) presents the intensity spatial distribution of the spatial spectrum measured in the far-field ($k_x$, $k_y$), for various input intensities $I$.
A ring structure appears as $I$ increases. The shape is clearly not symmetric. As will be discussed later, this is due to the anisotropic response of the crystal. \\
First, let us focus on the behaviour along the $k_x$ axis which corresponds to the $c$-axis conjugate.
We plot the $k_x$-profiles on the Fig.~\ref{fig:profiles}(a).
To improve the signal-to-noise ratio, we integrate the signal over 40 pixels along the $k_y$ axis.
With increasing optical intensities, more and more rings appear, indicating an increase of the nonlinear phase shift.
Typical profiles at $I=9$ mW.cm$^{-2}$ and $I=49$ mW.cm$^{-2}$ are shown in the Fig.~\ref{fig:profiles}(b-c).
The gray points count twice the number of rings $N_{\rm ring}$.
The measurement of the nonlinear refractive index based on sSPM requires a precise analysis of the curvature of the central spot.
Indeed, as detailed in \cite{Shen1984, Deng2005, Ramirez2010}, the nonlinear phase shift $\Delta \Psi_0$ is $2\pi{}N_{\mathrm{rings}}$ for a bright (i.e. concave) central spot and $(2N_{\mathrm{rings}} - 1)\pi$ for the dark (i.e. convex) case.
Note that in both cases, the contribution to the nonlinear phase shift of the appearance of a new ring is $2\pi$~\cite{durbin1981}, if the curvature of the central spot remains unchanged.
For example, in Fig.~\ref{fig:profiles}(b), we have $N_{\rm ring} = 4$ and a convex curve at $k_x=0$ (see inset) which gives $\Delta \Psi_0 = 7 \pi$.
In the Fig.~\ref{fig:profiles}(c), we have $N_{\rm ring} = 1$, a concave curve at $k_x=0$ and thus $\Delta \Psi_0 = 2 \pi$.
Finally, one obtains the nonlinear index of refraction $\Delta n$ as a function of the nonlinear phase, $\Delta n = \Delta \Psi_0/(k_{0}L_{z})$, with $L_{z} = 20\,\mathrm{mm}$.

\begin{figure*}[t]
 \includegraphics{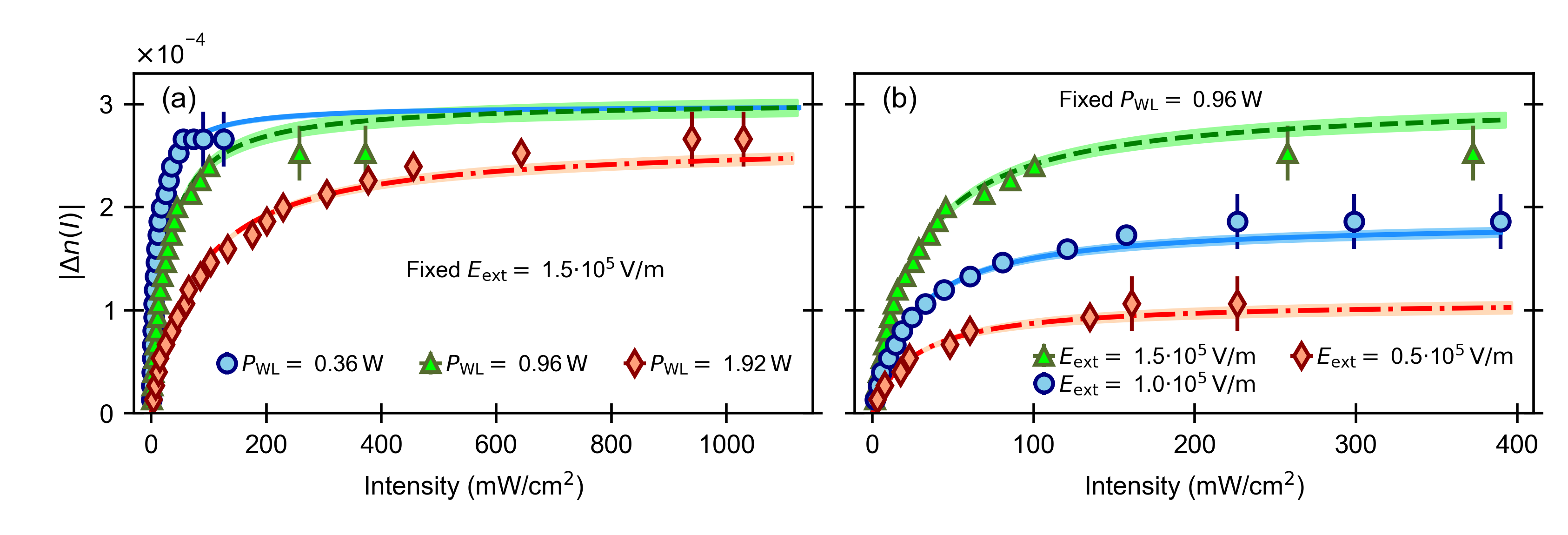}
 	\centering
    \caption{Nonlinear index of refraction along the $c$-axis for a given set of parameters of white light intensity and external electric field with respect to the incident optical intensity and corresponding fits using a isotopic saturable model (eq. (\ref{eq:Deltan_sat})).
    (a) Same applied external electric field of $1.5\times10^{5}\,\mathrm{V.m}^{-1}$ and
    different white light power $[0.36,\,0.96,\,1.92]\,\mathrm{W}$.
    (b) Same applied white light power (1.92 W), giving $I_{\mathrm{sat}} \simeq 25\,\mathrm{mW.cm}^2$ and different electric field $[0.5,\,1,\,1.5] \times10^{5}\,\mathrm{V.m}^{-1}$, from left to right.
    On both panels, vertical error bars are estimated looking at the cuts of the images and represent an uncertainty of one ring on the rings count at high intensity.
    Horizontal errors bars are smaller than the symbols.}
    \label{fig:fitNLI}
\end{figure*}

Moreover, the diameter of the rings saturates after a certain intensity.
As discussed above, such a saturation effect is expected for a photorefractive nonlinearity (in both isotropic and anisotropic descriptions).
Then, at very high optical intensity (above 300 mW.cm$^{-2}$), the power inside the central spot increases, reducing the contrast of the outer rings.
This observation can be explained considering Eq.~\eqref{eq:Deltan_sat}. In the regime where $I \gg I_{\rm sat}$, the nonlinear refractive index is independent of $I$, implying that the medium becomes homogeneous.
Therefore, no interference patterns can be observed in the far-field and one gets a clear Gaussian diffracted pattern.
This behavior has been also found in our simulations.
For a Kerr nonlinearity, the numerics did not reveal such a behavior.
This indicates that it is only related to the saturable nature of the PR nonlinearity.

From the rings counting procedure and the curvature analysis at $k_x=0$, the absolute value of $\Delta{}n$ is plotted in the Fig.~\ref{fig:fitNLI} versus the optical intensity $I$ of the incoming beam for different sets of external electric field and white light intensity.
In both configurations, we  observe an increase of $|\Delta{}n|$ for small $I$ and a clear saturation at high intensity.
Let us first consider the case of Fig.~\ref{fig:fitNLI}(a) where $E_{\text{ext}} = 1.5\times10^{5}\,\mathrm{V.m}^{-1}$ is fixed.
From left (blue-filled circles) to right (red diamonds), the white light power is increased from $0.36\,\mathrm{W}$ to $1.92\,\mathrm{W}$.
The values reached on the plateau, $\Delta n_{\text{max}}$, by the three curves are very close to each other.
Moreover when the white light intensity gets higher, the optical intensities needed to reach the plateau gets higher as well.
The data are fitted taking into account the isotropic model defined by eq.~(\ref{eq:Deltan_sat}) along the $c$-axis of the crystal.
The absolute variation of the nonlinear refractive index is then $|\Delta n(I)| = |\Delta n_{\text{max}}|\times\tilde I/(1 + \tilde I)$.
The three fitted curves are shown on Fig.~\ref{fig:fitNLI}(a) and the fitted parameters are presented in Tab.~\ref{tab:SummaryFit}.
The values extracted from the experimental data are in good agreement with the theoretical value, $|\Delta n_{\text{max}}| = 2.32\times10^{-4}$ given by Eq.~\eqref{eq:Deltan_sat}.

We present on Fig.~\ref{fig:fitNLI}(b) the case where the white light is set at $P_{\rm WL} = 1.9\,\mathrm{W}$ such as $\Isat$ is fixed to $25.7\,\mathrm{mW.cm}^{2}$ and different external electric fields $[0.5,\, 1,\,1.5] \times10^{5}\,\mathrm{V.m}^{-1}$, from left to right.
Here, it is clear that the plateau is reached for the same incident optical intensity, which is expected as the saturation intensity is fixed.
On the other hand, the maximum value obtained for $\Delta n_{\rm max}$ gets smaller as $\Eext$ decreases.
The extracted free parameters of the fits, $|\Delta n_{\rm max}|$ and $\Isat$, are summarised in Tab.~\ref{tab:SummaryFit}.

\begin{table}[b]
\caption{\label{tab:SummaryFit}
Summary of the fitted parameters extracted using an isotropic saturable model on the data
presented on Fig.~\ref{fig:fitNLI}(a) and Fig.~\ref{fig:fitNLI}(b).}
\begin{ruledtabular}
\begin{tabular}{cccc}
 & Symbols &$|\Delta n_{\text{max}}|\,(\times 10^{-4})$&
$I_{\text{sat}}\,(\mathrm{mW.cm}^{2})$\\
\hline
Same $E_{\text{ext}}$   & $\circ$ & $2.99 \pm 0.02$ & $9.0 \pm 0.2$\\
Fig.~\ref{fig:fitNLI}(a) & $\triangle$ & $3.03 \pm 0.08$ & $25.7 \pm 0.9$\\
                        & $\lozenge$ & $2.64 \pm 0.06$ & $74.7 \pm 3.3$\\ \hline
Same $I_{\text{sat}}$   & $\triangle$ & $3.03 \pm 0.08$ & $25.7 \pm 0.9$\\
Fig.~\ref{fig:fitNLI}(b) & $\circ$ & $1.86 \pm 0.05$ & $23.9 \pm 0.9$\\
                        & $\lozenge$ & $1.09 \pm 0.06$ & $24.9 \pm 2.2$\\
\end{tabular}
\end{ruledtabular}
\end{table}

As a summary, results presented in Figs. \ref{fig:fitNLI} (a) and (b) and in Tab. \ref{tab:SummaryFit} show that we have a control over the dynamics and asymptotic value of the nonlinear refractive index of the crystal with respect to the laser beam intensity. Indeed, we are able to measure an absolute value for the maximal nonlinear index of refraction in good agreement with the theory. Moreover, the estimation of the saturation intensity $I_{\text{sat}}$ seems to be accurate and grows, as predicted, with the white light intensity.
Those measurements thus suggest that the spatial self-phase modulation is a suitable tool to predict the behaviour of the photorefractive nonlinearity.

\begin{figure}[t]
 \includegraphics{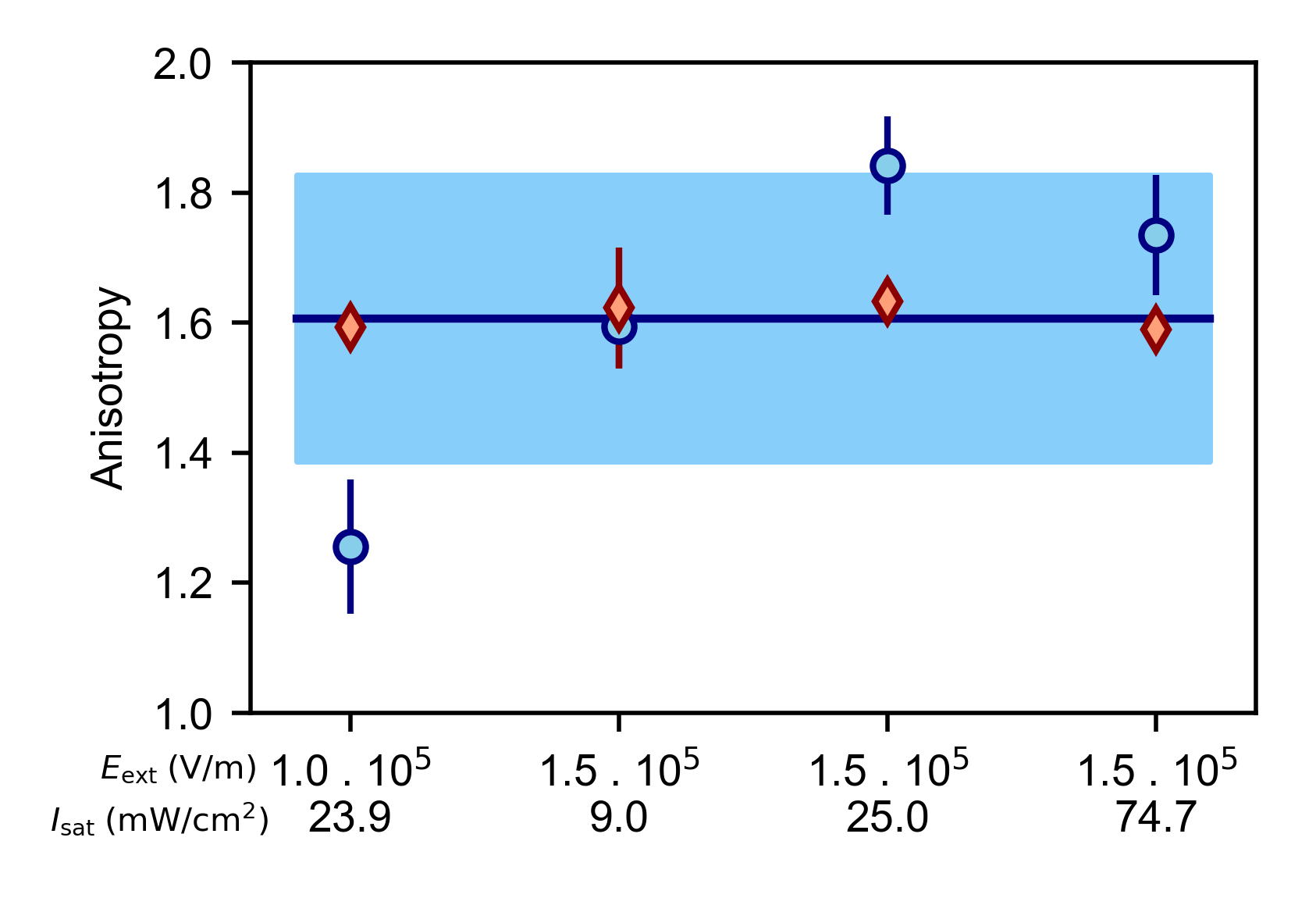}
    \caption{Measurement of the anisotropy of SBN.
    Blue circles: ratio of the fitted $|\Delta n_{\text{max}}|$ along the $k_x$ direction and the orthogonal direction, errorbars: uncertainty given by the fitting procedure.
    Blue solid line: mean value of the data, light-blue area: standard deviation of the data.
    Red diamonds: aspect ratios extracted directly from the images, averaged over the 10 images having the highest intensity.
    Error bars correspond to the standard deviation.}
    \label{fig:Aniso}
\end{figure}

As mentioned previously and shown on Fig.~\ref{fig:images}(a), the recorded images present a shape radically different from the expected circular rings.
To get better understanding on the observed anisotropy, we solved numerically the coupled Eqs.~(\ref{eq:NLS}) and~(\ref{eq:potential}).
The results shown in Fig.~\ref{fig:images}(b) for $I = 100$ mW.cm$^{-2}$ is, at least qualitatively, in very good accordance with the corresponding experimental image.
To compare with, we present in Fig.~\ref{fig:images}(c), the numerics obtained at the same input intensity in the isotropic approximation, solving directly Eq.~(\ref{eq:NLS_A}).
This allows us to clearly attribute the non-symmetric shape of the spectrum to the anisotropic response of the crystal, described by Eqs.~(\ref{eq:Deltan}) and (\ref{eq:potential}).\\
To be more quantitative, we apply the same experimental procedure on profiles obtained along $k_y$ to extract the nonlinear refractive index for each input intensity.
Although not shown here, a saturation is also observed.
We present on Fig.~\ref{fig:Aniso} the ratios of $|\Delta n_{\text{max}}|$ along $k_x$ over along $k_y$ extracted from the fits for all the different sets of parameters $E_{\text{ext}}$ and $I_{\text{sat}}$ (blue-filled circles).
Those values are compared to the aspect ratios extracted from the images (Fig.~\ref{fig:images}(a)), defined as the ratio between the diameter of the outer ring along the $k_x$-direction and along the perpendicular direction and averaged over the 10 images having the highest
intensity (red-filled diamonds).
We observe a very good agreement between the two methods used to estimate the anisotropy of the nonlinear crystal, and the value of 1.6 is coherent with the literature \cite{armijo2014}.
Moreover, our simulations indicate a slightly higher anisotropy of 2.0.
This measurement suggests that one can take benefit of the spatial self-phase modulation to access and characterise properly the anisotropy of a photorefractive crystal such as SBN.
If we want to describe entirely the diffracted pattern in the far-field, the anisotropic description of the PR effect is required.
However, if we are only interested in measuring the nonlinear refractive index of the medium along the $c$-axis ($k_x$ conjugated), the isotropic approximation is quantitatively pertinent.

\section{Conclusion}

To conclude, we have presented an absolute measurement of the nonlinear refractive index of a SBN photorefractive crystal by studying diffraction patterns in the far-field taking benefit of spatial self-phase modulation.
When a gaussian wavepacket is initially injected in a medium with isotropic nonlinear response, the expected signature of the sSPM consists in the appearance of new spatial frequencies in the far-field with the observations of concentric rings.
We have shown experimentally that for the photorefractive crystal, the spectral intensity distribution is not symmetric. 
By means of numerical simulations, we confirmed that such signal is due to the anisotropic response of the crystal.
We analysed the spatial spectrum and measured the nonlinear refractive index versus input optical intensity for various experimental conditions (external voltage and white light illumination).
Our measurements are in very good agreement with the isotropic model of the photorefractive effect when we consider the spectrum along $k_x$ which is the conjugated of the $c$-axis.
Finally, the anisotropy we measured is 1.6.

The proposed experimental technique is quick, easy to implement and might be applicable to any nonlinear crystals.
This is of great importance in the context of linear and nonlinear light transport investigations where precise calibration of the photoinduced nonlinear refractive index is required.

It's worth mentioning that we didn't discuss the transient regime of the photorefractive response which, in the present case, is of the order of few tens of seconds. 
During the transient, the refractive index goes from zero to its maximum value.
This regime opens new perspectives in probing the nonlinear light dynamics and will be the subject of future studies.

\section*{Acknowledgments}
The authors thank G. Labeyrie for fruitful discussions. 
This work has been supported by the the Region Sud, the French government, through the UCA$^{\rm JEDI}$ Investments in the
Future project managed by the National Research Agency (ANR) with the reference number ANR-15-IDEX-01 and the European Union's Horizon 2020 research and innovation programme under grant agreement No 820392 in the PhoQuS project framework.

\bibliography{refs.bib}

\end{document}